\documentclass[letter]{IEEEtran}
\usepackage{psfig,amssymb,amsmath,amsfonts,amsthm,graphicx}

\newtheorem{theorem}{\textbf{Theorem}}
\newtheorem{lemma}{\textbf{Lemma}}
\newtheorem{corollary}[theorem]{Corollary}

\title{Data Dissemination in Wireless Networks with Network Coding}
\author{Mohammad~Hamed~Firooz,~\IEEEmembership{Student~Member,~IEEE,}
        Sumit~Roy~\IEEEmembership{Fellow,~IEEE,}
\thanks{The authors are with the Department of Electrical Engineering, University of Washington, Seattle,
WA, 95195 USA e-mail: \{firooz,roy\}@uw.edu.}
}

\begin{document}
\maketitle
\begin{abstract}
We investigate the use of network coding for information dissemination over a wireless network. Using network coding allows for a simple, distributed and robust algorithm where nodes do not need any information from their neighbors. In this paper, we analyze the time needed to diffuse information throughout a network when network coding is implemented at all nodes. We then provide an upper bound for the dissemination time for ad-hoc networks with general topology. Moreover, we derive a relation between dissemination time and the size of the wireless network. It is shown that for a wireless network with $N$ nodes, the dissemination latency is between $O(N)$ and $O(N^2)$, depending on the reception probabilities of the nodes. These observations are validated by the simulation results.
\end{abstract}
\vspace{-2mm}
\section{Introduction}
The information dissemination problem, at its root, is a classical broadcast problem: sharing data residing at one node (source) with all others (destinations) in the network. Of late, a more modern version of the one-to-many (broadcast) problem has gained prominence; this is typically referred to as {\em data sharing} among multiple peer-to-peer (p2p) nodes, or the {\em all-to-all} problem. This problem arises when each node in a network obtains only {\em a
fraction of the total information} (e.g., part of a video-on-demand file or a software update) desired collectively by all. In a simplified version of the all-to-all data-dissemination problem, a source file desired by all is divided into $N$ mutually exclusive \emph{information packets}, and each packet is stored at exactly one node in the network \cite{kermarrec2003probabilistic, demers1987epidemic}. Every node's objective is to acquire the remaining $N-1$ pieces of the source file; the order in which each node receives the remaining information packets is not relevant.

Traditionally, the data dissemination problem over decentralized network architecture has focused on the impact of the  {\em  dissemination algorithm} designed to optimize a performance metric, such as dissemination latency (i.e., the time required for \emph{all nodes} to acquire the entire file \cite{minsky2002spreading,firoozICC}). Authors in \cite{deb2006algebraic,boyd2006randomized} showed that using Network Coding (NC) for dissemination in a wired network can improve dissemination latency.

Recently, authors in \cite{fragouli2008efficient} used NC to diffuse information in an ad-hoc wireless network. However, their analytical models have largely suffered from unrealistic assumptions that are unsuited to a wireless network--notably that of pure fail/success ($0-1$)--whereby each transmission is either successfully received by \emph{all} neighbors or fails. Our analysis advances the state of the art by using a more appropriate link model whereby, for each broadcast, sink nodes successfully receive the transmitted packet with a reception probability that is dependent upon the nodes' respective locations. We provide an upper bound for dissemination latency for a wireless network with general topology. Moreover, we show that in a connected wireless network, the dissemination latency, in the worst case, increases quadratically with the number of nodes in the network.

There is a plausible argument as to why network coding provides substantial benefits for data dissemination. In the beginning, each node has only a small fraction of the full file and seeks to gather the remaining pieces. With time, a node gathers some of the other pieces, but does not have any information regarding which pieces the neighboring nodes may possess.  At any instant, the profile of packets at any two nodes in the network will include a common and remaining non-overlapping subsets. Intuitively, this suggests that, if each node encodes all the data it presently contains via network coding and broadcasts it, recipient nodes will have acquired coded versions containing information about the missing pieces. After a sufficient number of such encoded packet transmissions from other nodes, each node will be able to decode the full file. Thereby, by using NC,  nodes do \emph{not} need extra information from other nodes concerning the state of the network \cite{xiumin2010data,firooz2012arxiv}.

In this paper, we focus only on dissemination latency and do \emph{not} consider the latency caused by encoding/decoding of NC, which has been studied separately in the literature \cite{wang2006practical}. In fact, authors in \cite{li2011rank} explore the design of a sparse network coding matrix that significantly decreases encoding/decoding time. Clearly, the net latency of data dissemination is the sum of our result and the encoding/decoding time.

The rest of the paper is organized as follows: Section \ref{S:system_model} describes the system model and basic assumptions used in this paper. The data dissemination using network coding in wireless networks is introduced in Section \ref{S:systemModel}. In Section \ref{S:stoppingTime}, we derive an upper bound on dissemination latency. Performance evaluations are presented in Section \ref{S:eval} and the paper concludes in Section \ref{S:conc}.

\emph{Notations}: Bold capitals (e.g. $\mathbf{A}$) represent matrices and bold lowercase symbols (e.g. $\mathbf{m}$) denote vectors. The $i$-th entry of a vector $\mathbf{m}$ is denoted by $m_i$ and superscript $^T$ denotes matrix transpose. $|S|$ represents the cardinality of a set $S$. For a set $S=\{\mathbf{x}_1,\ldots,\mathbf{x}_N\}$, the subspace spanned by elements of $S$ is called the subspace of $S$ and $dim(S)$ denotes the dimension of that subspace \cite{mceliece1987ffc}. The equality between two subspaces $S_1$ and $S_2$ is denoted by $S_1\equiv S_2$.
\vspace{-2mm}
\section{System Model}\label{S:system_model}

As usual, a network graph is denoted as $G(V,E)$, with $|V|=N$ nodes and links $E\subset V\times V$. We assume that the network is slotted (i.e., all nodes are synchronized) for simplicity and that all transmissions occur synchronously with a common clock. Further, without loss of generality, we assume that during each time slot, a node $v\in V$ can broadcast exactly one packet. When node $v$ broadcasts, node $u\in V$ receives the signal correctly with probability $P_{vu}$.

Clearly, in all broadcast wireless networks, the role of the multiple access or MAC protocol is fundamental to managing interference \cite{asterjadhi2010toward}. We consider an interference-free (orthogonal) access that allows only one node to transmit at a time. This includes, among others, a single-cell 802.11-type infrastructure network based on CSMA/CA
if all nodes lie within the (common) carrier sensing range \footnote{As is generally true, the carrier sensing range is larger than the transmission range.}
\cite{bianchi1996performance}. The probability of a node capturing the common channel at any time is assumed to be uniform among all nodes.

\vspace{-2mm}
\section{Data Dissemination using Network Coding}\label{S:systemModel} 
Assume that each node $u$ initially has a single \emph{information packet} $\mathbf{x}_u$ to be shared with every other node in the network. Hence, the set of unique (information) packets in the network, initially and at all subsequent times, is given by $ \{ \mathbf{x}_1 , \ldots , \mathbf{x}_N \}$ for a network with $N$ nodes. Each information packet is a vector of $r$ symbols, where each symbol is an element of a finite field $\mathbb{F}_{2^q}$, i.e., $\mathbf{x}_u\in \mathbb{F}^r_{2^q}$ for each node $u\in V$. For convenience, assume that $q$ divides the length of packets transmitted (otherwise, zero padding is applied). Moreover, all packets are linearly independent vectors in $\mathbb{F}_{2^q}$ \cite{mceliece1987ffc}, reflecting the fact that nodes have different information to share.
The results and derivations presented in this paper can be extended to a case when some nodes have more than one message and some have none or when all the messages are there with one particular node to start with.

With time, each node receives a sequence of linear combinations of information packets at the other nodes. Hence, after a sequence of broadcasts, node $u\in V$ possesses a set of coded messages, $S_u(t)$ at time (slot) $t$.
\begin{equation}
S_u(t) = \{\mathbf{m}_1, \mathbf{m}_2, ..., \mathbf{m}_{|S_u(t)|}\},
\end{equation}

Each message $\mathbf{m}_i$ is a linear combination of the underlying \emph{information packets}, initially possessed by the nodes and can be represented as
\begin{equation}
\mathbf{m}_i = \sum_{k=1}^{N}\alpha_{i,k}\mathbf{x}_k = \boldsymbol{\alpha}_i^T\mathbf{X}, \;\;\;i=1,2,...,|S_u(t)|,
\label{E:m=alphaX}
\end{equation}
where some of the coefficients $\alpha_{i,k}$ may be zero (if the corresponding information packet is not present at the transmitting node at that time). For each message $\mathbf{m}_i$, $\alpha_{i,k}$s are called its \emph{network coding coefficients}, and they are available through the header of the packet containing $\mathbf{m}_i$. As discussed in \cite{chou2003practical,firooz2010link}, in a network coding system, each packet consists of two parts: a header that contains the network coding coefficients and a body that carries the encoded message. This header is a price to pay to use the network coding. However, if the size of the information packets (and hence the size of the messages) is reasonably large, this overhead is negligible. That being said, for each message $\mathbf{m}_i$ at node $u$, network coding coefficients are available.

Clearly, $S_u(t)$ (set of messages at node $u$ at time $t$) spans a subspace in $\mathbb{F}^r_{2^q}$, as observed by rewriting Eq. \eqref{E:m=alphaX} in the following form:
\begin{equation}
\mathbf{M}_u(t) = \mathbf{A}_u(t)\, \mathbf{X},
\label{E:m=AX}
\end{equation}
where $\mathbf{A}_u$ is the coefficient matrix consisting of NC  coefficients and $\mathbf{X}$ contains the $N$ information packets in the network, given by
\begin{eqnarray*}
&&\mathbf{X} =[\mathbf{x}_1\;\mathbf{x}_2\ldots \mathbf{x}_N]^T, \mathbf{M}_u(t)= [\mathbf{m}_1\;\mathbf{m}_2\ldots \mathbf{m}_{|S_u(t)|}]^T,\\ \nonumber
&&\mathbf{A}_u(t)=
\left[
\begin{array}{cccc}
\alpha_{1,1}&\alpha_{1,2}&\ldots&\alpha_{1,N}\\
\vdots & \vdots & \ddots & \vdots \\
\alpha_{|S_u(t)|,1}&\alpha_{|S_u(t)|,2}&\ldots&\alpha_{|S_u(t)|,N}
\end{array}
\right].
\end{eqnarray*}
\vspace{-2mm}
\section{Stopping Time}\label{S:stoppingTime}
The data dissemination algorithm terminates when {\em all} nodes are able to decode the broadcast messages to recover the underlying $N$ set of information packets, which happens when Eq. \eqref{E:m=AX} for all $u\in V$ has a unique solution, i.e., when the coefficient matrix at each node has full rank $N$.

Matrix $\mathbf{A}_u$ has rank $N$ if and only if $S_u(t)$, the subspace scanned by messages in $u$ at time $t$, has dimension $N$. Hence, the stopping time $\mathcal{T}$ is defined as follows:
\begin{equation}
\mathcal{T}=\min_t\{dim(S_u(t))=N\,\,\forall u\in V\}.
\end{equation}

Clearly, $\mathcal{T}$ is an integer random variable over $[N,\infty)$\footnote{To diffuse $N$ packets, we need at least $N$ transmissions, which in wireless networks require at least $N$ time slots.}. We next seek the expected value $\mathbb{E}[\mathcal{T}]$ as a performance metric for algorithm design. In general, $\mathbb{E}[\mathcal{T}]$ is difficult to compute; hence, we resort to bounds.
\vspace{-3mm}
\subsection{Upper Bound for Mean Stopping Time}

By our formulation, every node initially starts with an information packet. In other words, at $t=0$, there is one and only one (independent) packet in $S_u(0)$, i.e., $dim(S_u(0))=1\,\,\forall u\in V$. With time, the information spreads to all nodes upon sharing via broadcast, resulting in a final per node dimension of $N$ at the time of stopping. Hence, each node dimension is raised by $N-1$ during the information dissemination, and the overall dimension increase among all the nodes is $N(N-1)$. Let us define $D(t)$ as the total dimension increase (among all the nodes) at time $t$. Obviously, $D(t)$ can be written as
$D(t)=\sum_{u\in V} dim(S_u(t))-N$.

Clearly, the information has spread to all nodes when $D(t)=N(N-1)$. Now, let $\mathcal{T}_i$ denote the number of time slots until the total dimension increases by $i(N-1)$. It can be written as $\mathcal{T}_i=\min_t\{D(t)\ge i(N-1)\}$.

\setlength{\textfloatsep}{4pt plus 1pt minus 1pt}
\begin{figure*}[t]
\begin{equation}\label{E:PS_u=S_v}
P(S_u(\mathcal{T}_i)\equiv S_v(\mathcal{T}_i)) \le
\sum_{k=1}^{N-1}
\min({i\over k},{{N-i}\over{N-k-1}})
\frac{\sum_{j=0}(-1)^j\!{{N-1}\choose j}\!{{N(i-j+1)-(i+2+k)} \choose {N-2}}}
{\sum_{j=0}(-1)^j{N\choose j}{{N(i-j+1)-(i+1)}\choose {N-1}}}.
\end{equation}
\vspace{-8mm}
\end{figure*}

By definition $\mathcal{T}_0=0$ and the information spreads to all the nodes at $\mathcal{T}_N$, i.e., $\mathcal{T}=\mathcal{T}_N$. The following lemma gives an upper bound for the probability of the message sets at two nodes spanning the same subspace when $t=\mathcal{T}_i$.
\begin{lemma}\label{Tm:P(S_u=S_v)}
At $t=\mathcal{T}_i$, the probability that two nodes (e.g., $u, v$) have the same subspace can be bounded by Eq. \eqref{E:PS_u=S_v}.
\end{lemma}
\emph{Proof}: See \cite{firooz2012arxiv}.

Finally, the following theorem gives an upper bound on the stopping time.
\begin{theorem} Let $\mathcal{T}$ be stopping time. Then
\begin{equation}\label{E:ET}
\mathbb{E}[\mathcal{T}] \!
\le\! {{2N\!(N\!\!-\!\!1)} \over {\sum\limits_{u,v\in V}P_{uv}}} \bigg(\! \sum_{i=1}^{N-1} \!{1\over 1\!\!-\!\!P\big(S_u(\mathcal{T}_i)\equiv S_v(\mathcal{T}_i)\big)} + N \bigg).
\end{equation}
\label{Tm:ET}
\end{theorem}
\emph{Proof}: See \cite{firooz2012arxiv}.

Since each node is initialized with a single packet, it needs to acquire the remaining $N-1$ packets from other nodes for the process to terminate. Hence a total of $N(N-1)$ successful packet transmissions must occur. Due to the broadcast nature of wireless, multiple receive nodes hear each transmission and may decode the transmitted packet (according to their reception probability; higher reception probability results in a higher chance of decoding). Therefore, the number of time slots required is inversely proportional to reception probability as captured by the first part of the upper bound. The second part of the upper bound represents the fact that a successfully received packet $v$ at a node is only useful if it does not belong to the subspace spanned by existing packets; i.e., it is `innovative'. Again intuitively, the probability of a packet being innovative at a node decreases with time, as the subspace spanned by existing packets is always monotonic non-decreasing.

In the following two corollaries, we consider two extreme cases, a fully connected wireless network and a sparsely connected network \cite{firooz2012arxiv}. These two cases illustrate how reception probability affects dissemination based on NC in a wireless network.


For any node $u\in V$, let define $V_u=\{v\in V|P_{uv}>0\}$. In other words, $V_u$ is set of nodes which are located in the transmission range of $u$\footnote{1) Note that $V_u$ contains node $u$ itself. 2) All the nodes still belong to one cell, i.e., they are all in sensing range of each other}. A wireless network $G(V,E)$ is considered fully connected when, for any node $u\in V$, $V_u=V$; i.e., every node is within the
transmission range of all the others.
\begin{corollary}
In a fully connected wireless network, the average stopping time is of order $O(N)$ when NC is applied.
\label{Tm:orderN}
\end{corollary}

Above, the corollary is consistent with the result of the data dissemination in a wired network when network coding is adapted. The authors in \cite{deb2006algebraic} show that the stopping time increases linearly with the size of the wired network.

A wireless network $G(V,E)$ is sparsely connected when the network is connected and for any $u\in V$, $\frac{|V_u|}{|V|} \ll 1$. In other words, in a sparsely connected network, there are only few nodes in transmission coverage of node $u$. An example of a sparsely connected wireless network is a linear network where each node can only communicate with its close neighbors.

\begin{corollary}
In a sparsely connected wireless network, the average stopping time is of order $O(N^2)$ when NC is applied.
\label{Tm:orderN2}
\end{corollary}

Clearly, Corollary \ref{Tm:orderN} gives the best achievable time, and Corollary \ref{Tm:orderN2} gives the worst time (largest number of time slots needed). In other words, the above corollaries show that for data dissemination in a wireless network with $N$ nodes, the average stopping time is between $O(N)$ and $O(N^2)$, independent of the underlying nodes reception probability and network topology.
\vspace{-2mm}
\section{Performance Evaluation}\label{S:eval}
In this section, we present results from a system simulation conducted using MATLAB R2008b that conforms to the data dissemination model described.

We assume that all nodes use the same transmission power $\mathcal{P}$ and QAM modulation with no channel coding to
broadcast packets. The wireless channel is assumed to be Rayleigh fading, and the path loss exponent is $\eta$. Assume node $u$ sends a packet to node $v$ which is $d(u,v)$ far away. Node $v$, the receiver, can decode successfully  the packet transmitted by $u$ if its received Signal-to-Noise (SNR) ratio exceeds a threshold
\begin{equation}
P_{u,v}=Pr({s_{u,v}\over{\mathcal{N}_0}}\ge z),
\end{equation}
where $s_{u,v}$ follows an exponential random variable with mean $\mathcal{P}\cdot d(u,v)^{-2}$ and the $\mathcal{N}_0$ is the variance of additive white Gaussian noise, assumed to be  $4\times 10^{-14}$ at all the receivers \footnote{Noise Power is calculated for the bandwidth of 10MHz and in temperature 300K. This value, however, does not affect the result of the simulation.}, and $z$ is the capture reception threshold whose value depends on channel coding and modulation. In our simulation, we set $z=45\text{dBm}$.

The results reported are based on simulations conducted for
two simple and useful topologies: regular linear and 2-D grid. Such structured topologies help with better understanding of the model behavior as the number of nodes increase.
\vspace{-3mm}
\subsection{Linear Grid}
\begin{figure}
\centering
\psfig{figure=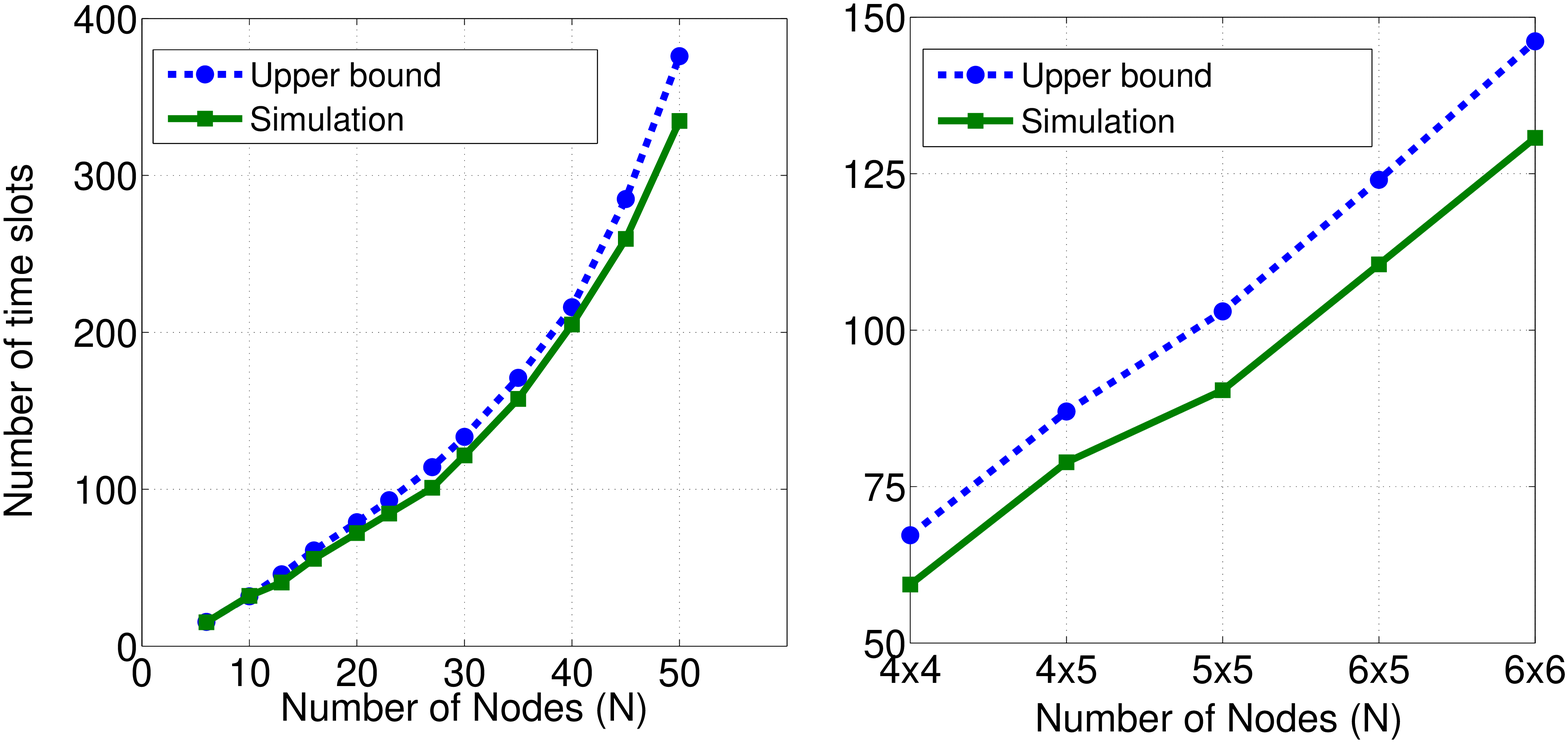,width=3.5in,height=2.0in}
\caption{Analytical and simulation results when size of network is changing
and transmission power is fixed (left) linear topology (right) grid topology. For both topologies we have $d=30$, $\mathcal{P}=20\times10^{-6}$, $\mathcal{N}_0=4\times10^{-14}$.}
\label{F:plot_linear_30}
\end{figure}

Here, nodes are located in a line, with equal distance, $d$, between neighbors. At first, we let the transmit power remain fixed and increase the size of the network by adding more nodes. Figure \ref{F:plot_linear_30} presents the simulation result and the analytical upper bound for the linear network. As one can see, the upper bound given in Eq. \eqref{E:ET} closely follows the trend of the simulation results.

When there are only a few nodes in the network, stopping time has a linear relation with the number of nodes in the network. However, when the size of the network keeps increasing, the linear relation is not valid anymore. This is consistent with our findings in Lemmas \ref{Tm:orderN} and \ref{Tm:orderN2}. For a small number of nodes in the network, nodes are in transmission range of each other; i.e., a transmitted packet is heard by all of the nodes in the network (with nonzero probability $P_{uv}>0$); hence, the stopping time is $O(N)$.  On the other hand, when the size of the network keeps expanding, after a while we have $P_{uv}=0$ for some nodes in the network and that affects the trend of the dissemination delay. In Figure \ref{F:plot_linear_30}, after $N=30$, the stopping time (from both the simulation and analytical result) starts to increase nonlinearly. In fact, one can see that the stopping time is $O(N^2)$ after $N=30$.

Finally, we change transmission power to see its effect on dissemination latency and the accuracy of the upper bound in Eq. \eqref{E:ET}. The result is presented in Figure \ref{F:plot_linear_P}. Clearly, decreasing transmission power reduces nodes' coverage and results in increased stopping time. However, the relation between the stopping time and the transmission power is very interesting and is sort of hidden in Eq. \eqref{E:ET}.

For a fixed network, we start with $0\text{dBm}$ power and decrease it to $-40\text{dBm}$. At first, nodes are in transmission range of each other and the relation between the transmission power and the stopping time is linear. However, after a point, nodes start falling out of the transmission range of each other. When that happens, stopping time starts increasing nonlinearly with transmission power. As one can see our formula in Eq. \eqref{E:ET} has the same trend as the simulation results.

To demonstrate the advantage of network coding, we compare the dissemination latency with network coding using computer simulation to a baseline, {\em random} non-NC algorithm. The non-NC approach - termed random selection - operates as follows: whenever a node captures the channel it {\em randomly selects}  an information message from its buffer and broadcasts the selected packet. Table \ref{T:NC_vs_baseline} compares the mean time needed to diffuse data using the two schemes \cite{firooz2012arxiv}.

\begin{table}
\centering
\caption{Stopping Time for dissemination algorithm in a linear network with and without network coding. $d=30$, $\mathcal{P}=20\times10^{-6}$}
\begin{tabular}{c|c|c|c|c}
\# nodes& 23 & 27 & 30 & 35
\\
\hline \hline
NC-based & 84.46 & 100.94 & 121.54 & 157.59
\\
\hline
random-selection&1189.6&2180.8&3148.9&3713.4
\\
\hline
\end{tabular}
\label{T:NC_vs_baseline}
\end{table}
\vspace{-3mm}
\subsection{2-D Grid}
In a 2-D grid topology, nodes are located on a equispaced 2-D lattice. 
As for the linear network, we first let the transmission power remain fixed while increasing the size of the network by adding more nodes. Figure \ref{F:plot_linear_30} presents the simulation result and the analytical upper bound.

\begin{figure}
\psfig{figure=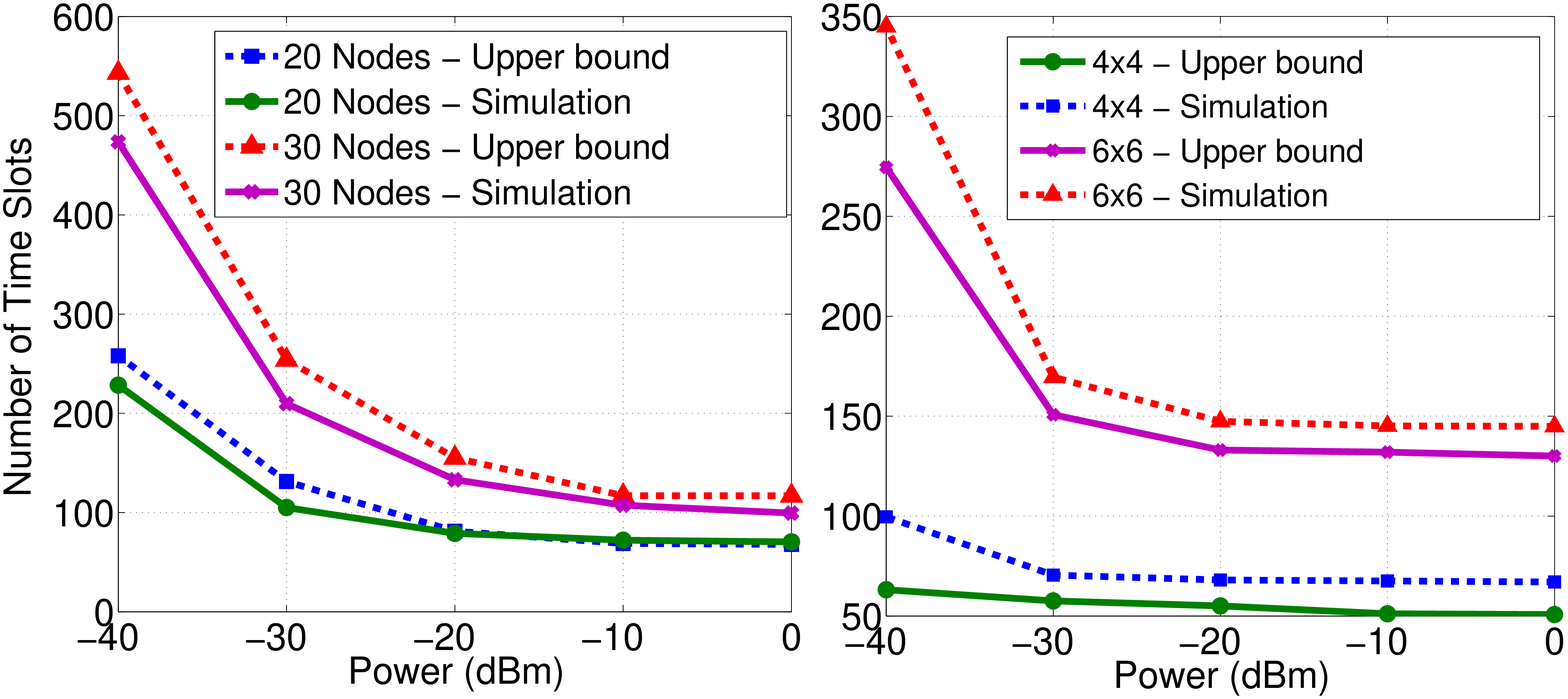,width=3.5in,height=2.0in}
\caption{Analytical upper bound and simulation results versus nodes' transmission power for (left) linear network (right) grid network. $\mathcal{N}_0=4\times10^{-14}$.}
\label{F:plot_linear_P}
\end{figure}

In an $m\times n$ grid network with equispaced $d$, the distance between every two nodes is less than or equal to $d\sqrt{m^2+n^2}$, which happens to be smaller than the transmission range of all the nodes in our simulation. In other words, for the fixed transmission power, each node can hear from all other nodes with nonzero probability. It is for this reason that the stopping time has a linear trend with the size of the network (Theorem \ref{Tm:orderN}). Finally, for different transmission power, analytical upper bounds and simulation results are presented in Figure \ref{F:plot_linear_P}.

\vspace{-2mm}
\section{Conclusion}\label{S:conc}
 In a wireless network with general topology, we provide an analytical upper bound for the amount of time needed to spread information through the whole network. Our result show that by using network coding the stopping time is between $O(N)$ and $O(N^2)$ where $N$ is number of nodes inside the network.

\vspace{-2mm}
\bibliographystyle{IEEEtran}
\bibliography{IEEEabrv,overall}

\end{document}